\documentclass[12pt]{iopart}

\newcommand{\al}{\alpha}

\usepackage{amssymb}
\usepackage{iopams}
\usepackage{footnote}

\begin{document}

\title[]{On energy exchange rate \\ and entropy production  %
operators \\ in quantum fluctuation-dissipation relations}

\author{Yu. E. Kuzovlev}

\address{Donetsk Institute for Physics and Technology %
of NASU \\  ul. R.\,Luxemburg 72, 83114 Donetsk, Ukraine}
\ead{kuzovlev@fti.dn.ua}

\begin{abstract}
For quantum systems with externally time-varied Hamiltonians %
a definition of operators of system's energy change rate (work per unit time) %
and entropy production observables is suggested and %
discussed in the context of rigorous statistical equalities %
(generalized fluctuation-dissipation relations) under the %
Jordan-symmetrized chronological operator ordering rule.
\end{abstract}


\vspace{2pc} \noindent{\it Keywords}\,: %
\,generalized fluctuation-dissipation relations, %
generalized fluctuation-dissipation theorems, %
fluctuation theorems, %
quantum fluctuation-dissipation relations, %
quantum fluctuation theorems

\section{Introduction}

The generalized fluctuation-dissipation relations (FDR) %
are statistical expression of such fundamental properies of %
microscopic dynamics as unitarity (phase volume conservation, %
in classical mechanics) and time reversibility. %
This rather old subject \cite{bk1,bk2,bk3} %
(see also \cite{fdr1,ufn1,z12} and references therein) %
came into ``second life'' during last %
fifteen years as ``fluctuation theorems'' or %
``fluctuation relations'' \cite{rmp,cht} (see also references %
there).

In their simplest forms, FDR deal with a Hamiltonian system %
whose Hamiltonian \,$\,H(x)$\, depends on some external %
parameters \,$x=x(t)$\, varying in a given way. %
A system may be very small, - e.g. a single atom under given %
external fields, - or arbitrary large, e.g. charge carrier in %
crystal lattice or other environment which can play role %
of ``thermal bath'', or thermostat.  %
Therefore temperature $T$ appears in FDR  %
generally as characteristics  of statistical ensembles %
although can be thought also as characteristics of %
system's own internal thermostat if any \cite{bk1}.

In respect to the parameters various systems can be %
sorted to ``closed'' and ``open'' \cite{bk2,bk3}. %
A closed system keeps its energy constant if external %
parameters are not changing, $x=$\,const\,.  %
In opposite, an open system can get unboundedly growing %
amount of energy even if parameters stay unchanging. %
Characteristic examples are the mentioned charge carrier %
(or probe particle of a fluid) %
\cite{bk10,i1,i2,dvr,tmf} and rotator in a fluid \cite{ufn1}  %
under non-zero external electric field (or force) and torque, %
respectively.  Instead of thermodynamical equilibrium, %
such systems at $x=$\,const\,$\neq 0$ become driven to (quasi-) %
steady non-equilibrium states. Qualitative difference  %
between these two sorts of systems naturally implies differences in %
formulation of FDR for them \cite{ufn1,z12} %
\footnote{\, %
In literature the word ``open'' usually refers to systems %
which contact with an outstanding thermostats (``environments''). %
We used it in this sense in several works on ``stochastic %
representation of deterministic interactions'' only  %
\cite{sr1,sr2,sr3,sr4,sr5,sr67}  %
but everywhere else, from \cite{bk1,bk2} to \cite{ufn1}, %
clearly used it in the above mentioned sense %
presuming that thermostat is included into system.
}\,. %

The aforesaid equally concerns classical and quantum %
theories. In quantum case, however, one meets %
specific problems involved by possible %
non-commutativity of variables (observables). %
For instance, in place of classical statistical identity %
 \cite{bk1}  %
\begin{eqnarray}
\langle\, e^{-E(t)/T}\,\rangle_0\,=\, 1\,\, \label{cr}
\end{eqnarray}
for change $E(t)=H_0(\Gamma(t))-H_0(\Gamma)$  %
of internal  energy of a system, %
 $H_0(\Gamma)=$ $H(\Gamma,x=0)$\, \cite{bk1,bk3,ufn1}, %
in quantum theory one has \cite{bk1}   %
\begin{eqnarray}
1\,=\,\langle\, e^{-H_0(t)/T}\, %
e^{\,H_0/T}\,\rangle_0\,=\, %
\langle\, e^{-[H_0\, +E(t)]/T}\, %
e^{\,H_0/T}\,\rangle_0\,=\, %
\label{qr}\\ =\,
\langle\, \overrightarrow{\exp} \, %
[\,-\int_0^1 e^{-\al H_0/T}\,E(t)\, %
e^{\,\al H_0/T}\,d\al\,/T\,]\,\,\rangle_0\, \nonumber
\end{eqnarray}
or, equivalently,
\begin{eqnarray}
1\,=\, \langle\,e^{\,H_0/T}\, e^{-H_0(t)/T}\, %
\rangle_0\,=\, %
\langle\, e^{\,H_0/T}\, e^{-[H_0 \,+E(t)]/T}\, %
\rangle_0\,=\, \label{qr1}\\  =\, %
\langle\, \overleftarrow{\exp} \, %
[\, -\int_0^1 e^{\,\al H_0/T}\,E(t)\, %
e^{-\al H_0/T}\,d\al\,/T\,]\,\rangle_0\,\,, \, \nonumber
\end{eqnarray}
where $H_0(t)=\, U^\dagger_t\,H_0\,U_t$\, %
is the internal energy's operator $H_0$ in the Heisenberg %
representation, with  %
\[
U_t\,=\,U(t,0)\,\,, \,\,\,\,\, %
U(t,t_0)\,=\, \overleftarrow{\exp} \, %
\left[ -\frac i\hbar \int^t_{t_0} %
H(x(\tau))\,d\tau\,\right]\,\,,
\]
being unitary operator of system's evolution in its Hilbert space, %
 $E(t)=H_0(t)-H_0$\,,  %
and $\langle\dots\rangle_0$  means averaging over %
canonical initial (at time $t=0$) %
density matrix (probability distribution) %
\begin{eqnarray}
\rho_{0}\,=\, \exp{[(F_0-H_0)/T]}\,\, \label{eq0}
\end{eqnarray}
(with ``free energy'' $F_0$ determined by the %
normalization condition), that is %
   $\langle\dots\rangle_0 =$ $\Tr\,\dots\,\rho_0$\,. %
These formulae clearly show that operator %
 $E(t)=H_0(t)-H_0$   can not be interpreted %
as operator of {\it \,observable\,} physical quantity, - %
``change of system's internal energy'' in mind, - %
since otherwise quantum analogue of Eq.\ref{cr} would  %
look as  $\Tr\, \exp{[-E(t)/T]}\,\rho_0\,=1$\,. %
The same can be said about ``change of system's full energy'' %
 $W(t)=H(t,x(t))-H(x(0))$ %
(analogue of classical %
  $W(t)=H(\Gamma(t),x(t))-H(\Gamma,x(0))$\,) with  %
  $H(t,x)\equiv U^\dagger_t\,H(x)\,U_t$\,. %
If so, then, correspondingly, operators %
\begin{eqnarray}
\frac {dW(t)}{dt} = \frac {dH(t,x(t))}{dt} \,=\, %
%\frac {dx(t)}{dt}\cdot \frac{\pa H(t,x(t))} {\pa x(t)} = %
\,U^\dagger_t\,  %
\frac {dH(x(t))}{dt}\,U_t\,\,, \,\label{dw}\\ %
\frac {dE(t)}{dt} =\, \frac {dH_0(t)}{dt} = %
U^\dagger_t\,\frac i\hbar\, %
[\,H(x(t))\,,\,H_0\,]\, U_t\,\, \,\label{de}
\end{eqnarray}
can not pretend to delegation of such observables  %
as ``rate of change'' of system's full energy or internal energy.

Significance of this difficulty was %
highlighted and discussed by H\"{a}nngi, Campisi and %
Talkner \cite{cht} (see there references to their %
earlier works) and later by other authors \cite{rmp}. %

Clearly, an appropriate solution of %
related problems, - first of all, formal definition of %
operator of system's ``energy change rate'' observable, - %
is impossible without definite choice of operator %
ordering rule when calculating quantum statistical %
moments and constructing quantum characteristic and %
probabilistic functionals.
We in the present paper consider solution of this %
problem in frame of one undoubtedly meaningful %
ordering rule which already was exploited by us  %
\cite{fdr1,sr1,sr2,sr3} as well as by many others %
(see e.g. references in \cite{fdr1}).

\section{Preliminary discussion of the problem  %
and its unconventional solution}

Let $\dot{E}(t)$ be Schr\"{o}dinger operator of the %
``internal energy change rate'' (IECR) to be defined, %
  $U^\dagger_t\,\dot{E}(t)\,U_t$  %
its Heisenberg form, and  $\mathcal{O}$ symbolizes some ordering of %
operator products and exponentials composed of %
\,{\it Heisenberg}\, operators. %
Then, at given $\mathcal{O}$\,, we want %
to find such generally non-zero Hermitian operator $\dot{E}(t)$ that
\begin{eqnarray}
\langle\, \exp\,\{\,-\frac 1T \int_0^t %
\dot{E}(\tau)\, d\tau\,\}\,\rangle_0\, %
\equiv\, \, \label{oqr}\\ \equiv\, %
\Tr\,\mathcal{O}\left\{\, \exp{[-\frac 1T %
\int_0^t U^\dagger_\tau\dot{E}(\tau) %
\,U_\tau\,d\tau\,]}\,\right\}\, %
\rho_0\,\,=\,1\, \, \, \nonumber
\end{eqnarray}
regardless of time variations of the %
Hamiltonian parameters $x(t)$\,. %
Existence of such operator $\dot{E}(t)$ gives rights to %
treat Eq.\ref{oqr} as true quantum analogue of classical %
Eq.\ref{cr} and at once reformulation of quantum %
identity (\ref{qr})-(\ref{qr1}) in terms of IECR. %

Here and below variables inside angle brackets, -  %
as $\dot{E}(t)$ in Eq.\ref{oqr}, - represent effective %
commutative (\,{\it c}-number valued) images of %
\,{\it Heisenberg\,} operator variables under the tracing, - %
as $U^\dagger_t\,\dot{E}(t)\,U_t$  %
in Eq.\ref{oqr}. Therefore time argument of the first %
reflects complete ``double'' time %
dependency of the second.

Notice that the ordering in Eq.\ref{oqr} does not touch  %
the initial density matrix, %
in accordance with standard postulate %
of quantum statistical mechanics. %
Interestingly, if we violate this postulate and introduce %
ordering of arbitrary operators  by rule \cite{bk3} %
\begin{eqnarray}
\langle\, \exp{[\int_0^t \Phi(\tau) %
\,d\tau\,]}\,\rangle\,\equiv\, %
\Tr\,\exp\,[\,\ln\,\rho_0\,+ %
\int_0^t U^\dagger_\tau\Phi(\tau) %
\,U_\tau\,d\tau\,]\,\,, \,\label{sd}
\end{eqnarray}
then Eq.\ref{oqr} is satisfied just %
at\, $U^\dagger_t\, \dot{E}(t)\,U_t\,=$ %
 $dE(t)/dt$\,, %
that is operators $E(t)$ and and (\ref{de}) %
appear to be really the ``internal energy change'' %
and IECR  observables. %
Other advantage of such unconventional rule   %
was underlined in \cite{bk3,fdr1}: under it all %
quantum FDR in terms of the angle brackets %
look exactly as classical ones.

\section{Jordan-symmetrized chronological ordering}

Obviously, the rule (\ref{sd}) has serious defect: %
it ignores time arguments of operators  %
under ordering and therefore can not be connected %
to some differential evolution equation.
This defect disappears if operator ordering %
is defined (quite conventionally) by
\begin{eqnarray}
\langle\, \exp{[\int_0^t \Phi(\tau)\, %
d\tau\,]}\,\rangle\,\equiv\, %
\,\, \label{cd}\\ \equiv\, %
\Tr\,\,\overleftarrow{\exp}\,[\, %
\frac 12 \int_0^t U^\dagger_\tau\Phi(\tau) %
\,U_\tau\,d\tau\,]\, %
\,\rho_{in}\,\, %
\overrightarrow{\exp}\,[\, %
\frac 12 \int_0^t U^\dagger_\tau\Phi(\tau) %
\,U_\tau\, d\tau\,]\,\, \nonumber
\end{eqnarray}
with $\rho_{in}=\rho_0$ or other initial density matrix. %

This definition involves natural chronological time %
ordering of observations: the later is one, %
the farther is its operator apart from initial density  %
matrix, in pleasant agreement with the causality principle. %
Therefore quantum characteristic functionals (CF) like (\ref{cd}) %
can be introduced via differential equations. Namely, %
if $\rho =\rho(t)$ is solution to equation
\begin{eqnarray}
\frac {d\rho}{dt}\,=\,\frac 12 %
\,\{\,U^\dagger_t\Phi(t)U_t\,\rho + %
\rho\,U^\dagger_t\Phi(t)U_t\,\}\,\, \, \label{hfe}
\end{eqnarray}
in Heisenberg representation or, equivalently,
\begin{eqnarray}
\frac {d\rho}{dt}\,=\,\frac 12 %
\,\{\,\Phi(t)\,\rho +\rho\,\Phi(t)\,\}\,+\, %
\frac i\hbar\, [\,\rho,H(x(t))]\,\,\, \label{qfe}
\end{eqnarray}
in Schr\"{o}dinger representation, %
with initial condition $\rho(t=0)=\rho_{in}$\,, %
then $\Tr\,\rho(t)$ coincides with CF (\ref{cd}), %
\begin{eqnarray}
\langle\, \exp{[\int_0^t \Phi(\tau) %
\, d\tau\,]}\,\rangle\,=\, %
\Tr\,\rho(t)\,\, \label{qf}
\end{eqnarray}

What is important, Eqs.\ref{qfe}-\ref{qf} give %
in fact simplest generalization of %
classical theory arising from it after replacement of %
the classical Liouville operator by quantum  %
Liouvile (von Neumann) super-operator %
while usual \,{\it c}-number product %
of variables and distributions (functions of phase point $\Gamma$) %
by symmetrized Jordan product of operators,\, %
  $A\circ B\equiv (AB+BA)/2$\, (for details ee e.g. \cite{fdr1}). %

For our purpose it is convenient to introduce %
three super-operators as follow:
\begin{eqnarray}
\mathcal{C}_A\,B\,\equiv\, [A,B]\,\,, \,\,\,\,\,  %
\mathcal{L}_A\,B\,\equiv\, %
-\frac i\hbar\, [A,B]\,\,, \,
\,\,\,\, % \nonumber\\
\mathcal{J}_A\,B\,\equiv\, A\circ B\,\,,
\end{eqnarray}
and rewrite Eq.\ref{qfe} in more abstract form,
\begin{eqnarray}
\frac {d\rho}{dt}\,=\, %
\{\,\mathcal{J}_{\Phi(t)}\,+\, %
\mathcal{L}_{H(x(t))}\,\}\, \rho\,\, \, \label{sqfe}
\end{eqnarray}
Following the same ordering rule, we can write also

\begin{eqnarray}
\langle\, B(t)\,\exp{[\int_0^t %
\Phi(\tau)\, d\tau\,]}\,A(0)\,\rangle %
\,=\, \, \nonumber\\ =\, %
\Tr\,\mathcal{J}_{B(t)} \,\, %
\overleftarrow{\exp}\,\{\int_0^t [\,\mathcal{J}_ %
{\Phi(\tau)}\,+ \mathcal{L}_{H(x(\tau))}\,]\,d\tau\,\}\, %
\,\mathcal{J}_{A(0)}\, \, \rho_{in}\,=\, %
\nonumber\\ \,=\, \label{sqf}  %
\Tr\, \mathcal{J}_{B(t)}\,\rho(t)\,=\, %
\Tr\, B(t)\,\rho(t)\,\,, \,
\end{eqnarray}
with arbitrary operators $A(t)$ and $B(t)$ and, %
clearly,  $\rho(t)$ again satisfying Eq.\ref{sqfe} but now with %
initial condition $\rho(t=0)= A(0)\circ\rho_{in}$\, %
\footnote{\, %
At that, of course, inside the angle brackets $B(t)$ %
and $A(t)$ also mean classical images of similar Heisenberg operator %
variables inside the trace, e.g.   %
 $U^\dagger_t\,B(t)\,U_t$\,.}\,.

\section{Internal energy change rate operator. %
Conventional solution of the problem}

\subsection{Simple derivation of the IECR operator}

Taking into account that
\[
\overleftarrow{\exp}\,[\int_0^t %
\mathcal{L}_{H(x(\tau))}\, d\tau\,]\,\,A\, %
=\, U_t\,A\,U^\dagger_t\,\,
\]
for any $A$\,, %
let us rewrite identity (\ref{qr})-(\ref{qr1}) %
in the spirit of chosen ordering rule:
\begin{eqnarray}
1\,=\,\Tr\, e^{-H_0/T}\, %
\overleftarrow{\exp}\,[\int_0^t %
\mathcal{L}_{H(x(\tau))}\,d\tau\,]\,\, %
(\,e^{\,H_0/T}\circ \rho_0\,)\,\, \label{jqr0}
\end{eqnarray}
On the other hand, if we dispose of IECR operator %
  $\dot{E}(t)$ then, according to Eqs.\ref{oqr} %
and \ref{cd}, the average (\ref{jqr0}) %
must be representable also in the form
\begin{eqnarray}
1\,=\,\Tr\,\,\overleftarrow{\exp}\,\{\int_0^t %
[\,-\frac 1T \,\mathcal{J}_{\dot{E}(\tau)}\,+  %
\mathcal{L}_{H(x(\tau))}\,]\, d\tau\,\}\, %
\,\rho_0\,\,=\,\, \label{jqr}\\
=\, \langle\, \exp\,\{\,-\frac 1T \int_0^t %
\dot{E}(\tau)\, d\tau\,\}\,\rangle_0\,\,
\, \label{qcr}
\end{eqnarray}
Moreover, if operators under the tracings in %
in Eqs.\ref{jqr0} and\ref{jqr} describe one and same %
observation process then they should coincide one with %
another at any time dependence of Hamiltonian parameters $x(t)$\,. %
Obviously, in Eq.(\ref{jqr0}) this is merely $\rho_0$\,. %
Hence, the IECR operator satisfies equality
\begin{eqnarray}
\rho_0\,=\,\overleftarrow{\exp}\,\{\int_0^t %
[\,-\frac 1T \,\mathcal{J}_{\dot{E}(\tau)}\,+  %
\mathcal{L}_{H(x(\tau))}\,]\, d\tau\,\}\, %
\rho_0\,\, \, \label{ie}
\end{eqnarray}
It in turn says that $\rho_0$ is ``eigenfunction'' %
of the super-operator contained in squire brackets, %
\begin{eqnarray}
[\,-\frac 1T \,\mathcal{J}_{\dot{E}(t)}\,+  %
\mathcal{L}_{H(x(t))}\,]\, \rho_0\,=\,0\, %
\, \, \label{ie1}
\end{eqnarray}

In order to resolve this operator equation,  %
we have to define operation of inversion of the Jordan product %
  $C=A\circ B$\,. This can be made at least if one %
of the Jordan multipliers is strictly positive or strictly %
negative, e.g. $B>0$\,. Then $A$ is expressable by
\begin{eqnarray}
A\,=\,\mathcal{J}_B^{-1}\,C \,\equiv\, %
\int_0^\infty e^{-\al B/2}\,C\, e^{-\al B/2}\, %
d\al\,\, \,\,\,\, (B>0)\,\,\, \label{ij}
\end{eqnarray}
It is easy to verify that this expression  indeed %
satisfies $A\circ B =C$\, if all $B$\,'s eigen-values %
are positive. This is just the case for $\rho_0$\,, by its %
very definition (\ref{eq0}).
Thus, solution of Eq.\ref{ie1} is
\begin{eqnarray}
\dot{E}(t)\,=\, \mathcal{J}_{\rho_0}^{-1}\, %
\,\frac {iT}\hbar \,[\rho_0,H(x(t))]\,= \,\,  %
\label{ed}\\ =\, %
\,\frac {iT}\hbar %
\int_0^\infty e^{-\al \rho_0/2}\, %
[\rho_0,H(x(t))]\,e^{-\al \rho_0/2}\, %
d\al \,\, \, \nonumber
\end{eqnarray}
This is the desired IECR operator.

%!!!

In order to see its relation to the %
 $E(t)=U^\dagger_t H_0U_t -H_0$\,, %
notice that
\begin{eqnarray}
\overleftarrow{\exp}\,\{\int_0^t %
[\,-\frac 1T \,\mathcal{J}_{\dot{E}(\tau)}\,+  %
\mathcal{L}_{H(x(\tau))}\,]\, d\tau\,\}\, %
A\,=\, \,\, \, \nonumber\\
=\, U_t\, %
\,\overleftarrow{\exp}\,[\,-\frac 1{2T}\int_0^t %
U^\dagger_\tau \dot{E}(\tau) U_\tau\,]\, d\tau\,]\, %
\,A\,\, %
\overrightarrow{\exp}\,[\,-\frac 1{2T}\int_0^t %
U^\dagger_\tau \dot{E}(\tau)U_\tau\,]\, d\tau\,]\, %
\,U^\dagger_t\,\, \nonumber
\end{eqnarray}
for any $A$\,. Therefore Eq.\ref{ie} %
can be transformed to  %
\begin{eqnarray}
U^\dagger_t\,e^{-H_0/T}\,U_t\,=\, %
e^{-[H_0+E(t)]/T} =\, \label{iec}\\ =\, %
\overleftarrow{\exp}\,[\,-\frac 1{2T}\int_0^t %
U^\dagger_\tau \dot{E}(\tau)U_\tau\,]\, d\tau\,]\, %
\,e^{-H_0/T}\,\,
\overrightarrow{\exp}\,[\,-\frac 1{2T}\int_0^t %
U^\dagger_\tau \dot{E}(\tau)U_\tau\,]\, d\tau\,]\,
\, \nonumber
\end{eqnarray}
Evidently, right-hand side of this equality %
is just Jordan-symmetrized chronological version %
of its left side. %

\subsection{Comparison with older results}

The above deduced expression (\ref{ed}) %
for the IECR observable operator coincides with result obtained,  %
in context of ``FDR for continuous quantum measurements'', %
in \cite{fdr1}. To show this, %
let us introduce, as there and in \cite{bk1,bk3,ufn1}, %
operator $-h(x)$ of system's interaction with %
external sources of its parameter variations (``external %
work sources''), so that $h(x=0)=0$ and
\[
H(x)=H_0-h(x)\,\,, \,\,\,\,\,\, %
[\rho_0,H(x)] = - [\rho_0,h(x)]\,\, \,
\]
One of reasons for use of $h(x)$ is %
that in many applications and models %
 $h(x)$ in fact involves much less number of system's %
degrees of freedom then total one (for other related reasonings %
see \cite{ufn1}). %
Besides, consider operators $h(x)$\,, %
 $h(t,x(t))=$ $U^\dagger_t\,h(x(t))\,U_t$\,, %
 $\dot{E}(t)$\,, $\dot{E}(t,t)\equiv $ %
 $U^\dagger_t\,\dot{E}(t)\,U_t$\,, etc.,  %
in basis of eigenstates of the ``unperturbed'' Hamiltonian %
 $H_0$\,, satisfying $H_0|\nu\rangle =E_\nu|\nu\rangle$\,. %
Then expression (\ref{ed}) takes form
\begin{eqnarray}
\dot{E}(t)\,= - %
\sum_{\mu,\,\nu}\, \frac {2iT}\hbar \cdot %
\frac {\rho_{0\,\mu}- \rho_{0\,\nu}}  %
{\rho_{0\,\mu}+ \rho_{0\,\nu}}\,\, %
h_{\mu\nu}(x(t))\,X_{\mu\nu}\, %
=\, \nonumber\\ =\, %
\sum_{\mu,\,\nu}\, \frac {2iT}\hbar \, %
\tanh{\left( \frac {E_\mu-E_\nu}{2T}\right)}\,  %
h_{\mu\nu}(x(t))\,\,X_{\mu\nu}\,\,, \,\label{ed1}
\end{eqnarray}

where\, $\rho_{0\,\mu}= %
\exp{[(F_0-E_\mu)/T]}$\,,\,  $X_{\mu\nu} = |\mu\rangle\langle\nu |$ %
and\, $h_{\mu\nu}(x)\,=\, \langle\mu |\,h(x)\,|\nu\rangle$\,. %
Substitution of this form to Eq.\ref{qcr} yields %
formula  visibly (accurate to designations) %
equivalent to formula (27) from \cite{fdr1} which was found %
by different method under the same ordering rule. %

Thus, Eqs.\ref{qcr} plus \ref{ed} %
and formula (27) from \cite{fdr1} give the same %
quantum generalization of classical statistical %
equality (\ref{cr}) in terms of continuously measured %
(observed) time-local quantum variable, $\dot{E}(t)$\,. %

Comparison between IECR $\dot{E}(t)$ (\ref{ed}) and %
``naive'' expression  $i[H(x(t)),H_0]/\hbar$\, %
corresponding to (\ref{de}), that is between (\ref{ed1}) %
and $(i/\hbar)\sum (E_\mu-E_\nu) %
 h_{\mu\nu}(x(t)) X_{\mu\nu}$\,, shows that the first %
(``true'') differs from the second (``naive'' %
\footnote{\, %
Recall, however, that the second is %
true in the framework of completely symmetrized ordering defined
by Eq.\ref{sd} \cite{bk3}. %
}\,) %
by factor \cite{fdr1}
\begin{eqnarray}
\Delta_{\mu\nu}\, =\, %
\frac {2T}{E_\mu-E_\nu} \, %
\tanh{\left( \frac {E_\mu-E_\nu}{2T}\right)}\,  %
\, \,\label{del}
\end{eqnarray}
suppressing contribution of high-frequency quantum transitions.  %

\section{Full energy change rate operator}

Now, consider changes of system's full energy,  %
  $W(t)=H(t,x(t))-H(x(0))$ %
(\,$H(t,x(t))=U^\dagger_t H(x(t))U_t$\,), and construct operator %
of time-local ``full energy change rate'' (FECR) observable %
which will be denoted as $\dot{W}(t)$\,. %
Again, we start from statistical identity \cite{rmp,cht}
\begin{eqnarray}
\langle\, e^{-H(t,x(t))/T}\, %
e^{\,H(x(0))/T}\,\rangle_{x(0)}\,=\, %
\label{qrj}\\ =\, %
\langle\,e^{\,H(x(0))/T}\, e^{-H(t,x(t))/T}\, %
\rangle_{x(0)}\,=\,e^{-\Delta F(t)/T}\,\, \,\nonumber
\end{eqnarray}
representing quantum variant of the
classical Jarzynski equality \cite{jar1,lpp}
\begin{eqnarray}
\langle\,e^{-W(t)/T}\, \rangle_{x(0)}\,=\, %
e^{-\Delta F(t)/T}\,\, \label{cje}
\end{eqnarray}
Here $\langle\dots \rangle_{x}$ means averaging over %
normalized canonical density matrix (probability distribution)
\begin{eqnarray}
\rho_{eq}(x)\,=\, \exp\, \frac %
{F(x)-H(x)}T\,\,, \,\label{eq}
\end{eqnarray}
and $\Delta F(t)=F(x(t))-F(x(0))$\,.  %

%!!!

As in previous section, first, rewrite the ``raw'' %
equality (\ref{qrj}) in terms of the Jordan-symmetrized %
chronological ordering:
\begin{eqnarray}
e^{-\Delta F(t)/T}\,=\,\Tr\,\{\, %
e^{-H(x(t))/T}\, %
\overleftarrow{\exp}\,[\int_0^t %
\mathcal{L}_{H(x(\tau))}\,d\tau\,]\, %
\,\times\, \nonumber\\ \times\, %
(\,e^{\,H(x(0))/T} \circ %
\rho_{eq}(x(0))\,)\,\}\,\, \label{jqrj0}
\end{eqnarray}
On the other hand, if there exists FECR operator %
  $\dot{W}(t)$ then this average equally can be  %
represented by
\begin{eqnarray}
e^{-\Delta F(t)/T}\,=\,\Tr\,\,\overleftarrow{\exp}\,\{\int_0^t %
[\,-\frac 1T \,\mathcal{J}_{\dot{W}(\tau)}\,+  %
\mathcal{L}_{H(x(\tau))}\,]\, d\tau\,\}\, %
\,\rho_{eq}(x(0))\,\,=\,\,\, \label{jqrj}\\
=\, \langle\, \exp\,\{\,-\frac 1T \int_0^t %
\dot{W}(\tau)\, d\tau\,\}\,\rangle_{x(0)}\,\,
\, \label{qcrj}
\end{eqnarray}
At that, again not only results of averaging (tracing) %
in (\ref{jqrj0}) and (\ref{jqrj}) must be identical at %
arbitrary trajectory $x(t)$ but the whole expressions %
under averaging too. Obviously, this means that %
the FECR operator is such that
\begin{eqnarray}
e^{F(x(0))/T}\,e^{-H(x(t))/T}\, %
=\, \label{fe}\\ \,=\, %
\overleftarrow{\exp}\,\{\int_0^t %
[\,-\frac 1T \,\mathcal{J}_{\dot{W}(\tau)}\,+  %
\mathcal{L}_{H(x(\tau))}\,]\, d\tau\,\}\, %
\rho_{eq}(x(0))\,\, \,\, \,\,\,\, \nonumber
\end{eqnarray}
Differentiation of this equality in respect %
to time leads to equation
\begin{eqnarray}
\frac d{dt}\,e^{-H(x(t))/T}\, =\, %
-\,\frac 1T \,\dot{W}(t)\circ e^{-H(x(t))/T}\,\,
\, \, \label{fe1}
\end{eqnarray}
just determining $\dot{W}(t)$ (we took into account %
that $\mathcal{L}_{H(x)}\, %
\exp{[-H(x)/T]} =0$\,). %
Formal solution to this equation is
\begin{eqnarray}
\dot{W}(t)\,=\, -\,T\, %
\mathcal{J}^{-1}_{\exp{[-H(x(t))/T]}}\, %
\left\{ \,\frac {d\,}{dt}\, %
e^{-H(x(t))/T} \right\} \,\,  \label{fd}
\end{eqnarray}
with super-operator $\mathcal{J}^{-1}_{\dots}$  %
defined by (\ref{ij}). %
Then, relation like (\ref{iec}) takes place,
\begin{eqnarray}
U^\dagger_t\,e^{-H(x(t))/T}\,U_t\,=\, %
e^{-\,[\,H(x(0))\,+\,W(t)\,]/T} =\, %
\,\,\,\,\,\,\,\, \label{fec} %
\end{eqnarray}
\[
=\,\overleftarrow{\exp}\, %
[\,-\frac 1{2T}\int_0^t %
U^\dagger_\tau \dot{W}(\tau)U_\tau\,]\, d\tau\,]\, %
\,e^{-H(x(0))/T}\,\,
\overrightarrow{\exp}\,[\,-\frac 1{2T}\int_0^t %
U^\dagger_\tau \dot{W}(\tau)U_\tau\,]\, d\tau\,]\,  %
\, \,\,\,\,\, \nonumber
\]

This result can be formulated more compactly %
if we introduce operator
\begin{eqnarray}
\Delta\dot{S}(t)\,=\,\frac 1T\,[\, %
\dot{W}(t)\,-\,\frac {dF(x(t))}{dt}\, %
]\,\,  \, \label{ds}
\end{eqnarray}
and rewrite Eqs.\ref{fe} and \ref{fe1} as follow,
\begin{eqnarray}
\rho_{eq}(x(t)) \, %
=\,\overleftarrow{\exp}\,\{\int_0^t %
[\,- \,\mathcal{J}_{\Delta\dot{S}(\tau)}\,+  %
\mathcal{L}_{H(x(\tau))}\,]\, d\tau\,\}\, %
\rho_{eq}(x(0))\,\,, \, \label{fe0}\\
\frac {d\rho_{eq}(x(t))}{dt}\,=\, %
-\, \Delta\dot{S}(t)\circ \rho_{eq}(x(t))\,
\, \, \label{fe2}
\end{eqnarray}
Consequently, Eq.\ref{fd} takes form
\begin{eqnarray}
\Delta\dot{S}(t)\,=\, -\, %
\mathcal{J}^{-1}_{\rho_{eq}(x(t))}\, %
\,\frac {d\rho_{eq}(x(t))}{dt}\,=\,  %
\label{fd0}\\ =\,
-\int_0^\infty e^{-\al \rho_{eq}(x(t))/2}\, %
\frac {d\rho_{eq}(x(t))}{dt}\, %
e^{-\al \rho_{eq}(x(t))/2}\, d\al \,\, \nonumber
\end{eqnarray}
Thus, the desired FICR operator is found.

%!!!

\section{Discussion and resume}

We have constructed (Hermitian) operators of ``full energy %
change rate'' (FECR) $\dot{W}t)$\, and ``internal energy %
change rate'' (IECR) $\dot{E}(t)$\,  such that with their %
help quantum analoguess \cite{rmp,cht} of the classical %
Jarzynski equality \cite{jar1,lpp}  and Bochkov-Kuzovlev %
equality \cite{bk1,lpp} can be formulated in terms of continuously %
measured  quantum variables (observables).  %
In parallel we introduced operator  $\Delta\dot{S}(t)$ (\ref{ds}) %
which in many applications may delegate time-local %
entropy production observable, - at least in closed systems %
(see Introduction) with ``positional parameters'' (see %
\cite{ufn1}), -  or more generally (for closed systems) energy  %
absorption (desorption) per unit time %
which includes both irreversibly %
dissipated and revertible parts. In case of open systems %
(or some closed systems with ``force parameters'' \cite{ufn1}), %
however, the entropy production operator should be %
redefined, instead of (\ref{ds}), by %
 $\Delta\dot{S}(t)=\dot{E}(t)/T$\,. %

All that objects were constructed in the framework %
of Jordan-symmetrized chronological operator ordering %
rule (see Sec.3). It in fact envelopes also usual %
``two-point'' form of the mentioned statistical equalities,
Eq.\ref{qr} and Eq.\ref{qrj}, %
where total change of system's energy %
during all the observation time $t$ is thought %
as difference of results of two instant measurements %
of the energy at initial and final time %
moments \cite{rmp,cht}. Formally, this is not worse, %
and even better, recipe than performing %
(infinitely) many measurements %
of the IECR or FECR. From physical point of view, however,  such %
``two-point'' recipe seems too fantastic if a system %
under consideration is macroscopically large. If so, then %
it is more reasonable to try to integrate data from %
many measurements of small-time energy changes, each %
conducted through a small part of total number of %
system's degrees of freedom. %

Anyway it is interesting whether the ``continuous'' formulation %
of above considered statistical equalities can be extended to %
more general generating fluctuation-dissipation relations %
which exploit, in addition to unitarity of quantum evolution, %
also its time reversibility. Such attempt was made in \cite{fdr1} %
but it leaved a lot of questions not quite clear. We hope, however, %
that all ``blanks'' will be properly filled in a not far future.

%\begin{eqnarray}
%
%\end{eqnarray}

\,\,\,

---------------------------

\,\,\,

\end{document}